\documentclass[preprintnumbers,article,amsmath,amssymb,floatfix,10pt,prd,onecolumn,
superscriptaddress,nofootinbib]{revtex4-2}
\usepackage{bm}
\usepackage{amsfonts}
\usepackage{latexsym}
\usepackage[latin1]{inputenc}
\usepackage{graphicx}
\usepackage{amsmath}
\usepackage{palatino}
\usepackage{mathpazo}
\usepackage{textcomp}
\usepackage{float}
\usepackage{booktabs}
\usepackage{dcolumn}
\usepackage{ragged2e}
\usepackage{hyperref}
\hypersetup{colorlinks,citecolor=blue}
\hypersetup{colorlinks=true,linkcolor=red,filecolor=magenta,    urlcolor=blue}
\usepackage{amsmath}
\usepackage{xcolor}
\usepackage{orcidlink}
\usepackage{epsfig}
\usepackage[caption=false]{subfig}
\usepackage{commath}

\usepackage{multirow}
\newcommand{\be}{\begin{equation}}
\newcommand{\ee}{\end{equation}}
\newcommand{\bea}{\begin{eqnarray}}
\newcommand{\eea}{\end{eqnarray}}
\newcommand{\eeas}{\end{eqnarray*}}
\newcommand{\beas}{\begin{eqnarray*}}
\newcommand{\m}{\mathring}

\def\jnl@style{\it}
\def\aaref@jnl#1{{\jnl@style#1}}

\def\aaref@jnl#1{{\jnl@style#1}}

\def\aj{\aaref@jnl{AJ}}                   % Astronomical Journal
\def\apj{\aaref@jnl{ApJ}}                 % Astrophysical Journal
\def\apjl{\aaref@jnl{ApJ}}                % Astrophysical Journal, Letters
\def\apjs{\aaref@jnl{ApJS}}               % Astrophysical Journal, Supplement
\def\apss{\aaref@jnl{Ap\&SS}}             % Astrophysics and Space Science
\def\aap{\aaref@jnl{A\&A}}                % Astronomy and Astrophysics
\def\aapr{\aaref@jnl{A\&A~Rev.}}          % Astronomy and Astrophysics Reviews
\def\aaps{\aaref@jnl{A\&AS}}              % Astronomy and Astrophysics, Supplement
\def\mnras{\aaref@jnl{Mon.~Not.~Roy.~Astron.~Soc.}}             % Monthly Notices of the RAS
\def\prd{\aaref@jnl{Phys.~Rev.~D}}        % Physical Review D
\def\prc{\aaref@jnl{Phys.~Rev.~C}}  % Physical Review C
\def\prl{\aaref@jnl{Phys.~Rev.~Lett.}}    % Physical Review Letters
\def\qjras{\aaref@jnl{QJRAS}}             % Quarterly Journal of the RAS
\def\skytel{\aaref@jnl{S\&T}}             % Sky and Telescope
\def\ssr{\aaref@jnl{Space~Sci.~Rev.}}     % Space Science Reviews
\def\zap{\aaref@jnl{ZAp}}                 % Zeitschrift fuer Astrophysik
\def\nat{\aaref@jnl{Nature}}              % Nature
\def\aplett{\aaref@jnl{Astrophys.~Lett.}} % Astrophysics Letters
\def\apspr{\aaref@jnl{Astrophys.~Space~Phys.~Res.}} % Astrophysics Space Physics Research
\def\physrep{\aaref@jnl{Phys.~Rep.}}      % Physics Reports
\def\physscr{\aaref@jnl{Phys.~Scr}}       % Physica Scripta
\def\commat{\aaref@jnl{Comm.~Math.~Phys.}}              % Communications in Mathematical Physics
\def\science{\aaref@jnl{Science}}               % Science
\def\cqg{\aaref@jnl{Classical Quant.~Grav.}}            % Classical and Quantum Gravity
\def\jpcs{\aaref@jnl{JPCS}}                                     % Journal of Physics Conference Series
\def\ijmpd{\aaref@jnl{Int.~J.~Mod.~Phys.~D}}                    % International Journal of Modern Physics D
\def\grg{\aaref@jnl{Gen.~Relat.~Gravit.}}               % General Relativity and Gravitation
\def\rpp{\aaref@jnl{Rep.~Prog.~Phys.}}          % Reports on Progress in Physics
\def\npa{\aaref@jnl{Nucl.~Phys.~A}}        % Nuclear Physics A
\def\lrr{\aaref@jnl{Living Rev.~Rel.}}                   % Living reviews in relativity
\def\jcap{\aaref@jnl{J.~Cosmology Astropart.~Phys.}}    % Journal of cosmology and astroparticle physics
\def\rmp{\aaref@jnl{Rev.~Mod.~Phys.}}   %Reviews of modern physics
\def\epjc{\aaref@jnl{Eur.~Phys.~J.~C}} 
\def\plb{\aaref@jnl{~Phy.~Lett.~B}} 
\def\mpla{\aaref@jnl{Mod.~Phy.~Lett.~A}} 
\def\arxiv{\aaref@jnl{arxiv.org}}

\allowdisplaybreaks[1]
\begin{document}
%\color{red}
\color{black}       %% For one column
\title{Isotropization of locally rotationally symmetric Bianchi-I universe in $f(Q)$-gravity}
%\end{document}
\author{Avik De\orcidlink{0000-0001-6475-3085}}
\email{avikde@utar.edu.my}
\affiliation{Department of Mathematical and Actuarial Sciences, Universiti Tunku Abdul Rahman,\\
Jalan Sungai Long,
43000 Cheras,
Malaysia}

\author{Sanjay Mandal\orcidlink{0000-0003-2570-2335}}
\email{sanjaymandal960@gmail.com}
\affiliation{Department of Mathematics, Birla Institute of Technology and
Science-Pilani,\\ Hyderabad Campus, Hyderabad-500078, India.}

\author{J.T. Beh \orcidlink{0000-0001-9597-3099}}
\email{bjtheng@hotmail.com}
\affiliation{Institute of Mathematical Sciences,
Universiti Malaya,
50603 Kuala Lumpur,
Malaysia}

\author{Tee-How Loo\orcidlink{0000-0003-4099-9843}}
\email{looth@um.edu.my}
\affiliation{Institute of Mathematical Sciences,
Universiti Malaya,
50603 Kuala Lumpur,
Malaysia}

\author{P.K. Sahoo\orcidlink{0000-0003-2130-8832}}
\email{pksahoo@hyderabad.bits-pilani.ac.in}
\affiliation{Department of Mathematics, Birla Institute of Technology and
Science-Pilani,\\ Hyderabad Campus, Hyderabad-500078, India.}
%
%%%%%%%%%%%%%%%%%%%%%%%%%%%%%%%%%%%%%  DATE  %%%%%%%%%%%%%%%%%%%%%%%%%%%%%%%%%%%%
\date{\today}
\begin{abstract}
Despite having the somewhat successful description of accelerated cosmology, the early evolution of the universe always challenges mankind. Our promising approach lies in a new class of symmetric teleparallel theory of gravity named $f(Q)$, where the non-metricity scalar $Q$ is responsible for the gravitational interaction, which may resolve some of the issues. To study the early evolution of the universe, we presume an anisotropic locally rotationally symmetric (LRS) Bianchi-I spacetime and derive the motion equations. We discuss the profiles of energy density, equation of state and  skewness parameter and observe that our models archive anisotropic spatial geometry in the early phase of the universe with a possible presence of anisotropic fluid and as time goes on, even in the presence of an anisotropic fluid, the universe could approach isotropy due to inflation and the anisotropy of the fluid fades away at the same time.
\end{abstract}

\maketitle

\date{\today}
\section{Introduction}\label{sec1}

Even though the current Universe is believed to be essentially isotropic and homogeneous, without any special point or direction, and given by the Friedmann-Lema\^itre-Robertson-Walker (FLRW) metric, it may not have been so at its beginning near the initial singularity, nor does it necessarily have to be so in the future. In recent times the Wilkinson Microwave Anisotropy Probe (WMAP) dataset \cite{wmap,wmap1,wmap2} requires some additional morphology than the standard isotropic and homogeneous model of the universe. The observationally supported inflationary paradigm has the remarkable property that it isotropizes the baby universe into today's FLRW geometry. For a complete version one should still relax to the consideration of both spatial inhomogeneity and anisotropy and then study its evolution into the observed amount of homogeneity and isotropy. So as a first step one may consider Bianchi type cosmological models, which form a large and almost complete class of relativistic cosmological models, which are homogeneous but not necessarily isotropic, the other choice being the locally symmetric Kantowski-Sachs spacetime. Moreover, by studying an almost FLRW-like model with less symmetries, we can reasonably understand the isotropic ones, which can be considered their special sub-case \cite{pitrou}. In the present discussion we consider a special type of Bianchi universe, the locally rotationally symmetric (LRS) Bianchi type-I model to denote the anisotropic state of the universe, given by the metric in the Cartesian coordinates
\begin{equation} \label{metric}
ds^2 = -dt^2 + A^2(t) dx^2 + B^2(t) (dy^2 + dz^2).
\end{equation} 

All of our results can be extended to the Bianchi type I model without much effort. C.R. Fadragas et al. \cite{Fadragas/2014} explored a detailed dynamical analysis of anisotropic cosmologies with the presence of a scalar field. In particular, they studied Kantowski-Sachs, LRS Bianchi-I and Bianchi-III cases, and their outcomes are compatible with observations such as de Sitter, quintessence-like, or stiff-dark energy solutions. Moreover, in the last few decades, Bianchi cosmologies are seeking more interest in observational cosmology since the WMAP data \cite{wmap} suggest that the standard cosmological model with a positive cosmological constant resembles the Bianchi morphology \cite{Jaffe,Jaffe1,Jaffe2,Jaffe3,Jaffe4}. Also, these results indicate that the universe should have achieved a slightly anisotropic spatial geometry despite the inflation, which is contrary to generic inflationary models \cite{Guth,Guth1,Guth2,Guth3,Guth4,Guth5,Guth6}. 
Recently a wide range of Bianchi cosmology with the observational data have been studied (see details in \cite{A1,A2,A3,A4,A5}).

Furthermore, the accelerated expansion of the universe was confirmed by the Supernovae Cosmology Project in 1998, which used the data of IA Supernovae \cite{riess}. This gives rise to various theories in order to explain the acceleration. In general relativity (GR), the existence of an unknown form of energy in the universe, called the dark energy (DE), which possesses an exotic property such as negative pressure leads to a negative equation of state (EoS) parameter. To bypass the undetected DE, as an alternative to GR, the modified gravity theories were explored by modifying the Einstein-Hilbert action, but keeping the geometry intact. $f(R)$-theories of gravity are the simplest and most successful ones in this direction. Nevertheless, other kind of theories were also considered in the past by altering the underlying geometry but not disturbing the Lagrangian, such as the teleparallel theory equivalent to the GR (TEGR) and the symmetric teleparallel theory equivalent to the GR (STEGR). In these kind of theories, flat space is considered and the very special (symmetric and metric-compatible) Levi-Civita connection, used in GR, is replaced by an affine connection which has either non-vanishing torsion (in TEGR) or non-metricity (in STEGR) as the guiding force of gravity and its extension was proposed in \cite{Jimenez/2018}. The details about these three basic theories, we discussed in the below section.

A wide range of aspects have been studied in the context of $f(Q)$-gravity, such as its covariant formulation \cite{zhao}, spherically symmetric configuration \cite{lin}, energy conditions \cite{Mandal/2020b}, cosmography \cite{Mandal/2020c}, signature of $f(Q)$-gravity in cosmology \cite{signa}, as an alternative to $\Lambda$CDM theory \cite{lcdm} and also a special kind of $f(Q)$-gravity, called the Weyl type $f(Q)$-gravity \cite{xu,xu1}. 
The geodesic deviation equation in $f(Q)$-gravity was also studied and some fundamental results were obtained \cite{gde}. Recently, another array of very interesting results have been published in $f(Q)$-gravity (check details in the references herein \cite{Q1,Q2,Q3,Q4,Q5,Q6}). However, all the above literatures have focused to explore the present interests of the universe, by considering the isotropic and homogeneous FLRW metric in Cartesian coordinates, making sure of the coincident gauge, whereas in our present work we are aiming to study the early evolution of the universe in $f(Q)$-gravity with an anisotropic but homogeneous background metric, one significant leap in the cosmological application of this new gravity theory. 

In the standard theories of gravity governed by Einstein's field equations, the evolution from an anisotropic universe into an FLRW one, termed as the isotropization, can be achieved by a period of inflationary expansion, as discussed by \cite{wainwright}. Isotropization is a vital issue as it discusses whether the universe can result in isotropic solutions without the need of fine tuning the model parameters \cite{shamir,saridakis}. Naturally, for any new theories of gravity it is worthwhile to study the isotropization process and are considered to be one of the important milestones for that theory of gravity towards acceptance as an alternative to GR. It is no way sufficient but atleast it is necessary to revive the standard cosmological paradigms in any proposed alternative of GR. It is interesting to investigate whether these modified $f(R)$, $f(T)$ and $f(Q)$ theories can accommodate an anisotropic universe with or without the presence of anisotropic perfect fluid. Several studies were carried out in this direction for the first two theories, for example, in $f(R)$ theories anisotropic geometry was studied using the exponential and power-law volumetric expansions in \cite{anisotropyfR}. In \cite{anisotropyfR1}, detailed phase-space analysis of Kantowski-Sachs geometries was carried out, particularly for $f(R)=R^n$ and isotropization was achieved independent of the initial anisotropy. In the context of $f(T)$ theories, anisotropic background was considered by Rodrigues et al. in several occasions \cite{rod1,rod2,rod4,rod5}. Among others, models which can reproduce the early universe (assuming inflation) and the late-time accelerated expanding universe were obtained \cite{rod2}; it was shown that our universe might live a quintessence like state even if anisotropic models were considered \cite{rod5}; in \cite{rod4}, isotropization in a LRS-Bianchi background metric was considered and analysis showed that the nonlinear model, terms of higher order in $T$ are favored by observational data.  

The present article is organized as follows: on the geometrical view, we discus three fundamental theories of gravity in Sec. \ref{sec2a}. Sec. \ref{sec2} discusses the basic formalism of $f(Q)$-gravity. In Sec. \ref{sec3}, we derive the conservation equation for anisotropic spacetime. Then, we formulate equations of motion in the LRS-BI model in Sec. \ref{sec4}. In Sec. \ref{sec5}, we discuss anisotropic cosmological models with the choice of a linear $f(Q)$ function, which helps us to compare our results with GR. Sec. \ref{sec6} analyzes the cosmological models for modified $f(Q)$-gravity. Finally, gathering all of our results, we conclude in Sec. \ref{sec7}.

\section{Overview of geometrical trinity of gravity:}\label{sec2a}

The two  main building blocks of a spacetime are a metric tensor $g_{\mu\nu}$ and an affine structure, which is determined by a connection $\Gamma^{\alpha}_{\mu\nu}$ \cite{schr}. 
These two structures are completely independent in nature, but together helps to define the geometrical objects and that allow for conveniently classify geometries. 
The deviation of the connection from being metric is measured by the non-metricity
\begin{equation}
Q_{\alpha\mu\nu}\equiv \nabla_{\alpha}g_{\mu\nu},
\end{equation}
and its antisymmetric part defines the torsion
\begin{equation}
T^{\alpha}_{\mu\nu}=2\Gamma^{\alpha}_{[\mu\nu]}.
\end{equation}
Among all the possible connections, the Levi-Civita connection is the unique connection  that is 
symmetric and metric-compatible. Its components may be expressed by the Christoffel symbol
\begin{equation}
%\lbrace^{\alpha}_{\mu\nu}\rbrace
\mathring{\Gamma}^\lambda{}_{\mu\nu}=\frac{1}{2}g^{\alpha\lambda}(g_{\lambda\nu,\mu}+g_{\mu\lambda,\nu}-g_{\mu\nu,\lambda}).
\end{equation}
It is convenient to describe the general connection as
\begin{equation}
\Gamma^{\alpha}_{\mu\nu}=%\lbrace^{\alpha}_{\mu\nu}\rbrace
\mathring{\Gamma}^\lambda{}_{\mu\nu}+K^{\alpha}_{\mu\nu}+L^{\alpha}_{\mu\nu}.
\end{equation}
Notice that, the non-torsional part of the connection $\Gamma^{\alpha}_{\mu\nu}$ is the Levi-Civita connection, whereas contorsion and disformation have torsional transformation properties under change of coordinates. Further more, after gathering the relevant geometrical objects, we can use them to characterize a spacetime as follows:
\begin{itemize}
\item \textit{Metric}: the connection is metric-compatible, which indicates that $Q_{\alpha \mu\nu}(\Gamma, g)=0$. The length of vectors is conserved in metric spaces, hence non-metricity gauges how much their length changes when we parallel transport them.
\item \textit{Torsionless}: $T^{\alpha}_{\mu\nu}(\Gamma)=0$ and the connection is symmetric. The non-closure of the parallelogram generated when two infinitesimal vectors are parallel carried along one other is measured by torsion. As a result, it is commonly assumed that parallelograms do not close when torsion is present.
\item \textit{Flat}: $R^{\alpha}_{\beta\mu\nu}=0$ and the connection is not curved. Curvature is the rotation that a vector undergoes as it travels parallel along a closed curve. This creates a barrier for comparing vectors defined at various places in spacetime. However, in flat spaces, vectors do not rotate as they are conveyed, giving a stronger sense of parallelism at a distance. This is why theories developed in these environments are known to as teleparallel.
\end{itemize}
In Einstein's general relativity formulated on a metric and torsionless spacetime and attributed gravity to the curvature. However, it is natural to wonder, as Einstein did later, whether gravity may be attributed to the other qualities, such as torsion and non-metricity. 
	So far these three theories of gravity equivalently described GR and knocking into shape a geometrical trinity of gravity. 
	The usual formulation of GR, for example, assumes a Levi-Civita connection, which requires vanishing torsion and non-metricity, but its teleparallel equivalent (TEGR) assumes a Weitzenbock connection, which entails zero curvature and non-metricity \cite{maluf}. 
	The Weitzenbock condition of the vanishing of the sum of the curvature and torsion scalar was studied in a gravitational model in a Weyl-Cartan spacetime in \cite{Haghani}.
	Another similar formulation of GR, known as the symmetric teleparallel equivalent of GR, is a relatively unmapped field (STEGR). 
	The gravitational interaction is described by the non-metricity tensor Q, which takes into account vanishing curvature and torsion. 
	The STEGR was first presented in a brief paper \cite{nester}, in which the authors emphasize that the formulation brings a new perspective to GR, and that the gravitational interaction effects, via non-metricity, have a character similar to the Newtonian force and are derived from a potential, namely the metric. 
	The formulation, on the other hand, is geometric and covariant. 
	Therefore, to represent the same physical interpretation, GR can be described by %	Einstein-Hibert action as $\int \sqrt{-g}R$, 
	the integrand in Einstein-Hibert action as $R$,
	the integrand in action of teleparallel equivalent $T$ %$\int \sqrt{-g}T$ 
	\cite{olmo} and Coincident GR $Q$ %$\int \sqrt{-g}Q$ 
	\cite{heisenberg}. 
	The equivalent descriptions to GR by curvature, torsion and non-metricity provides the starting point to modified theories of gravity once the respective scalar replaced by the arbitrary functions. 
	Models of Ricci or torsion scalar general functions have already been widely investigated in the literature. 
	On general FLRW backgrounds, the cosmic realization of $f(R)$ theories forces them to stay near to GR, but models based on $f(T)$ suffer from substantial coupling issues \cite{golo}. 
	On generic FLRW backgrounds, the significant coupling difficulties that may be observed in $f(T)$ theories are absent in $f(Q)$ models. 
	Also, the predictions of the $f(Q)$ and $f(T)$ models correspond in the small-scale quasistatic limit, but that at higher scales the $f(Q)$ models generically transmit 2 scalar degrees of freedom that are absent in the case of $f(T)$.
	These two degrees of freedom vanish around maximally symmetric backgrounds, resulting in the strong coupling problem that has been explored \cite{jim}. 
	Moreover, one can see one of the interesting study \cite{fcai} where the formalism of $f(T)$-gravity was developed aligning with GR. 
	Although these three theories of gravity are considered to be completely equivalent, these two counterparts of GR almost remained unnoticed until recently when the extension in these two theories, respectively $f(T)$ and $f(Q)$ theories of gravity in line with the $f(R)$ extension of GR were investigated as an alternative of dark energy source and compared with the $f(R)$ theories \cite{fTfR}. 
	Among others, the second order field equations of these two theories, unlike the fourth order equations in the metric $f(R)$ theory were found to be advantageous. Furthermore, it is well-known that, theories of gravity on curvature and torsion are almost in their mature stage, but theories on the non-metricity is under development. 
	So, in this work, we are focusing to explore a new possibility to study of our universe in $f(Q)$-gravity.

\section{Basic Formalism of $f(Q)$-Gravity}
\label{sec2}
In $f(Q)$-gravity theory, the spacetime is constructed by using the symmetric teleparallelism and non-metricity condition, that is, $R^\rho{}_{\sigma\mu\nu} = 0$ and $Q_{\lambda\mu\nu} := \nabla_\lambda g_{\mu\nu} \neq 0$. The associated connection coefficient is given by 
\begin{equation} \label{connc}
\Gamma^\lambda{}_{\mu\nu} = \mathring{\Gamma}^\lambda{}_{\mu\nu} +L^\lambda{}_{\mu\nu} 
\end{equation}
where $\mathring{\Gamma}^\lambda{}_{\mu\nu}$ is the Levi-Civita connection and $L^\lambda{}_{\mu\nu}$ is the disformation tensor. This implies that 
\begin{equation*}
L^\lambda{}_{\mu\nu} = \frac{1}{2} (Q^\lambda{}_{\mu\nu} - Q_\mu{}^\lambda{}_\nu - Q_\nu{}^\lambda{}_\mu) \,.
\end{equation*}
In addition, we define the superpotential tensor
\begin{equation} \label{P}
P^\lambda{}_{\mu\nu} := \frac{1}{4} \left( -2 L^\lambda{}_{\mu\nu} + Q^\lambda g_{\mu\nu} - \tilde{Q}^\lambda g_{\mu\nu} -\frac{1}{2} \delta^\lambda_\mu Q_{\nu} - \frac{1}{2} \delta^\lambda_\nu Q_{\mu} \right) \,
\end{equation}
and using it, the non-metricity scalar 
\begin{equation} \label{Q}
Q = Q_{\lambda\mu\nu}P^{\lambda\mu\nu} = -\frac{1}{2}Q_{\lambda\mu\nu}L^{\lambda\mu\nu} + \frac{1}{4}Q_\lambda Q^\lambda - \frac{1}{2}Q_\lambda \tilde{Q}^\lambda \,. 
\end{equation}
The action of $f(Q)$-gravity is given by
\begin{equation*}
S = \int \left[\frac{1}{2\kappa}f(Q) + \mathcal{L}_M \right] \sqrt{-g}\,d^4 x
\end{equation*}
where $g$ is the determinant of the metric tensor and $\mathcal{L}$ is the matter Lagrangian. By varying the action with respect to the metric, we obtain
\begin{equation} \label{FE1}
\frac{2}{\sqrt{-g}} \nabla_\lambda (\sqrt{-g}f_QP^\lambda{}_{\mu\nu}) -\frac{1}{2}f g_{\mu\nu} + f_Q(P_{\nu\rho\sigma} Q_\mu{}^{\rho\sigma} -2P_{\rho\sigma\mu}Q^{\rho\sigma}{}_\nu) = \kappa T_{\mu\nu}.
\end{equation}
 Nevertheless, this equation is not in a tensor form, and it is only valid in the coincident gauge coordinate \cite{Jimenez/2018}. 

On the other hand, by using (\ref{connc}), we can have the following relations between the curvature tensors corresponding to $\Gamma$ and $\mathring{\Gamma}$:
\begin{equation}
R^\rho{}_{\sigma\mu\nu} = \mathring{R}^\rho{}_{\sigma\mu\nu} + \mathring{\nabla}_\mu L^\rho{}_{\nu\sigma} - \mathring{\nabla}_\nu L^\rho{}_{\mu\sigma} + L^\rho{}_{\mu\lambda}L^\lambda{}_{\nu\sigma} - L^\rho{}_{\nu\lambda} L^\lambda{}_{\mu\sigma}
\end{equation}
and so
\begin{align*}
R_{\sigma\nu} &= \mathring{R}_{\sigma\nu} + \frac{1}{2} \mathring{\nabla}_\nu  Q_\sigma + \mathring{\nabla}_\rho L^\rho{}_{\nu\sigma} -\frac{1}{2} Q_\lambda L^\lambda{}_{\nu\sigma} - L^\rho{}_{\nu\lambda}L^\lambda{}_{\rho\sigma} \nonumber \\
R &= \mathring{R} + \mathring{\nabla}_\lambda Q^\lambda - \mathring{\nabla}_\lambda \tilde{Q}^\lambda -\frac{1}{4}Q_\lambda Q^\lambda +\frac{1}{2} Q_\lambda \tilde{Q}^\lambda - L_{\rho\nu\lambda}L^{\lambda\rho\nu} \,.
\end{align*}
Therefore, by using the symmetric teleparallelism condition, we can rewrite the field equations in (\ref{FE1}) as
\begin{equation} \label{FE2}
f_Q \mathring{G}_{\mu\nu} + \frac{1}{2}g_{\mu\nu}(Qf_Q - f) + 2f_{QQ} \mathring{\nabla}_\lambda Q P^\lambda{}_{\mu\nu} = \kappa T_{\mu\nu}
\end{equation}
where $$\mathring{G}_{\mu\nu} = \mathring{R}_{\mu\nu} - \frac{1}{2} g_{\mu\nu} \mathring{R}$$ and $T_{\mu\nu}$ is the energy-momentum tensor.

To study non-trivial isotropization in the evolution process of the universe, once the anisotropic type spacetime metric is considered, the EoS parameter of the gravitational fluid should, in principle, also be generalized to exhibit an anisotropic character to give a more sensible model. With the isotropization of the universe, the fluid also isotropizes to display a vanishing skewness parameter and isotropic pressure. The energy-momentum tensor for the anisotropic fluid is defined as
\begin{equation} \label{T}
T^\mu_\nu = \text{diag}(-\rho, p_x, p_y, p_z) \,.
\end{equation}
where $\rho$ denotes the energy density of the fluid, $p_x$, $p_y$ and $p_z$ are the pressures along $x$, $y$ and $z$ coordinates which assume respective directional Equation of state (EoS) parameters $\omega_x$, $\omega_y$ and $\omega_z$. We parametrize the deviation from isotropy by setting $\omega_x = \omega$ and then denoting the deviations along $y$ and $z$ directions by the skewness parameter $\delta$, where $\omega$ and $\delta$ are possibly functions of time. Using the metric (\ref{metric}), the components of $T_{\mu\nu}$ are given by the following:
\begin{equation}
T_{00} = \rho \,, \quad  T_{11} = A^2 p_x \,, \quad T_{22} = B^2 p_y \,, \quad  T_{33} = B^2 p_z.
\end{equation}

%%%%%%%%%%%%%%%%%%%%%%%%%%%%%%%%%%%%%%%%%%%%%%%%%%%%%%%%%%%%%%%%%%%%%%%%%%%%
%%%%%%%%%%%%%%%%%%%%%%%%%%%%%%%%%%%%%%%%%%%%%%%%%%%%%%%%%%%%%%%%%%%%%%%%%%%%

\section{Conservation of energy-momentum}\label{sec3}
We open the present discussion with a significant result in symmetric teleparallelism, by proving that the divergence of the gravitational sector in the field equations of the modified $f(Q)$-gravity in the LRS-BI anisotropic universe vanishes. Thus we make sure that during the present discussion we do not have to impose any additional constraint on the function $f$ to restrict the movement of the test particles in a geodesic. This is the first proof of the null divergence of the energy-momentum tensor in the modified $f(Q)$-theory in anisotropic universe which is one of the basic principles of GR and staple for most of the gravity theories. 

First and foremost, we calculate the non-metricity scalar as
\begin{equation}\label{q}
Q = -2 \left( \frac{\dot{B}}{B} \right)^2 -4 \frac{\dot{A}}{A} \frac{\dot{B}}{B}.
\end{equation} 
Hence, $Q$ is only time-dependent, so we can write
\begin{equation*}
\nabla_\lambda Q = \nabla_0 Q = \partial_t Q =: \dot{Q}.
\end{equation*}

Taking divergence of (\ref{FE2}), we get
\begin{align}\label{div}
\m{\nabla}_\mu T^{\mu\nu} = &f_{QQ}\m{G}^{\mu\nu} \m{\nabla}_\mu Q + \frac{1}{2}f_{QQ} Q g^{\mu\nu} \m{\nabla}_\mu Q +2 f_{QQ} P^{\lambda\mu\nu} \m{\nabla}_\mu (\m{\nabla}_\lambda Q) \nonumber \\&+2f_{QQ} (\m{\nabla}_\lambda Q) (\m{\nabla}_\mu P^{\lambda\mu\nu}) + 2f_{QQQ} P^{\lambda\mu\nu} (\m{\nabla}_\mu Q)(\m{\nabla}_\lambda Q).
\end{align}
For the given metric (\ref{metric}) we can calculate [see Appendix]
\begin{align}
\left( \m{G}^{\mu\nu} +\frac{1}{2}Q g^{\mu\nu}\right) \m{\nabla}_\mu Q &= \dot{Q} \left[ 4\frac{\dot{A}}{A}\frac{\dot{B}}{B} +2\left( \frac{\dot{B}}{B}\right)^2 \right] \\
2 P^{\lambda\mu\nu} \m{\nabla}_\mu (\m{\nabla}_\lambda Q) &= \dot{Q}\left[\frac{1}{2} \left( \frac{\dot{A}}{A} \right)^2 -2\frac{\dot{A}}{A}\frac{\dot{B}}{B} \right] \\
2 (\m{\nabla}_\lambda Q)(\m{\nabla}_\mu P^{\lambda\mu\nu}) &=  \dot{Q} \left[ -\frac{1}{2} \left( \frac{\dot{A}}{A} \right)^2 -2\frac{\dot{A}}{A}\frac{\dot{B}}{B} -2\left( \frac{\dot{B}}{B} \right)^2 \right] \\
2 P^{\lambda\mu\nu} (\m{\nabla}_\mu Q)(\m{\nabla}_\lambda Q) &= 0. \label{d2}
\end{align}
Combining (\ref{div})-(\ref{d2}) we conclude
\be\m{\nabla}_\mu T^{\mu\nu}=0.\ee
For the fluid given by (\ref{T}), this results into
\be \label{ec}
\dot{\rho}+[3(1+\omega) H+2\delta H_y]\rho=0.
\ee
The presence of the term $\delta H_y\rho$ in (\ref{ec}) appeared due to the anisotropy of the fluid gurantees a non-constant energy density even if we consider the EoS $\omega=-1$ and vice-versa, unlike in the isotropic perfect fluid scenario for conventional vacuum energy.

%%%%%%%%%%%%%%%%%%%%%%%%%%%%%%%%%%%%%%%%%%%%%%%%%%%%%%%%%%%%%%%%%%%%%%%%%%%%
%%%%%%%%%%%%%%%%%%%%%%%%%%%%%%%%%%%%%%%%%%%%%%%%%%%%%%%%%%%%%%%%%%%%%%%%%%%%

\section{Equations of motion in the LRS-BI model}\label{sec4}

In this section, we derive the expressions of $\rho$, $p_x$, $p_y$, $p_z$, $\omega$, $\delta$. Using equation \eqref{metric} and \eqref{FE2}, we find the following equations of motion:
\begin{align}
 \rho &= \frac{f}{2} + f_Q \left[4 \frac{\dot{A}}{A} \frac{\dot{B}}{B} + 2 \left( \frac{\dot{B}}{B} \right)^2 \right] \\
 p_x &= -\frac{f}{2} + f_Q \left[-2 \frac{\dot{A}}{A} \frac{\dot{B}}{B} -2 \frac{\ddot{B}}{B} - 2\left( \frac{\dot{B}}{B} \right)^2 \right] -2\frac{\dot{B}}{B} \dot{Q} f_{QQ} \\
 p_y &= -\frac{f}{2} + f_Q \left[ -3 \frac{\dot{A}}{A} \frac{\dot{B}}{B} - \frac{\ddot{A}}{A} - \frac{\ddot{B}}{B} - \left(\frac{\dot{B}}{B} \right)^2 \right] - \left( \frac{\dot{A}}{A} +\frac{\dot{B}}{B} \right) \dot{Q} f_{QQ} \\
p_z &= p_y.
\end{align}
From the consideration of the anisotropic fluid
\begin{align*}
T^\mu_\nu &= \text{diag}(-1,\omega_x, \omega_y, \omega_z)\rho \\
&= \text{diag}(-1, \omega, (\omega+\delta), (\omega+\delta)) \rho,
\end{align*}
the equations of motion can be expressed as
\begin{align}
\rho &= \frac{f}{2} + f_Q \left[4 \frac{\dot{A}}{A} \frac{\dot{B}}{B} + 2 \left( \frac{\dot{B}}{B} \right)^2 \right] \label{qq1}\\
\omega\rho &= -\frac{f}{2} + f_Q \left[-2 \frac{\dot{A}}{A} \frac{\dot{B}}{B} -2 \frac{\ddot{B}}{B} - 2\left( \frac{\dot{B}}{B} \right)^2 \right] -2\frac{\dot{B}}{B} \dot{Q} f_{QQ} \label{qq2}\\
(\omega+\delta)\rho &= -\frac{f}{2} + f_Q \left[ -3 \frac{\dot{A}}{A} \frac{\dot{B}}{B} - \frac{\ddot{A}}{A} - \frac{\ddot{B}}{B} - \left(\frac{\dot{B}}{B} \right)^2 \right] - \left( \frac{\dot{A}}{A} +\frac{\dot{B}}{B} \right) \dot{Q} f_{QQ}.\label{qq3}
\end{align}

We have the directional Hubble parameters 
\be 
H_x=\frac{\dot{A}}{A},\quad H_y=\frac{\dot{B}}{B}, \quad H_z=\frac{\dot{B}}{B}
\ee
and the average Hubble parameter
\be\label{H}
H=\frac{1}{3}\frac{\dot{V}}{V}=\frac{1}{3}\left[\frac{\dot{A}}{A}+2\frac{\dot{B}}{B} \right]
\ee
where the spatial volume is
\be\label{V}
V=AB^2.
\ee
The rate of expansion is evaluated by anisotropy parameter
\be
\Delta=\frac{1}{3}\sum_{i=1}^3\left(\frac{H_i-H}{H}\right)^2=\frac{2}{9H^2}\label{delta}\left(H_x-H_y\right)^2.
\ee
Therefore, we also have
\be\label{36}
H_y^2+2H_xH_y=3H^2\left(1-\frac{\Delta}{2} \right).
\ee
The expansion scalar $\theta(t)$ and shear $\sigma(t)$ of the fluid are given by
\be \theta(t)=\frac{\dot{A}}{A}+2\frac{\dot{B}}{B},\quad \sigma(t)=\frac{1}{\sqrt{3}}\left( \frac{\dot{A}}{A}-\frac{\dot{B}}{B}\right).\ee 
First we express $\rho$, $\omega$ and $\delta$ in terms of the non-metricity scalar $Q$, the average Hubble parameters $H$ and directional Hubble parameters $H_x,H_y$. We note that $\frac{\partial}{\partial t}\left(\frac{\dot{A}}{A}\right)=\frac{\ddot{A}}{A}-\left(\frac{\dot{A}}{A}\right)^2$ and $Q=-2H_y^2-4H_xH_y$. 

From (\ref{qq1}) we get
\be \rho=\frac{f}{2}-Qf_Q.\label{qqq1}\ee
From (\ref{qq2}) we get
\be
\omega\rho=-\frac{f}{2}-2\frac{\partial}{\partial t}\left[H_yf_Q\right]-6Hf_QH_y\label{qqq2}
\ee
and from (\ref{qq3}),
\be (\omega+\delta)\rho=-\frac{f}{2}-\frac{\partial}{\partial t}\left[f_Q(H_x+H_y)\right]-3Hf_Q\left(H_x+H_y \right)\label{qqq3}.\ee
Therefore, (\ref{qqq1}) and (\ref{qqq2}) together imply
\be\omega=\frac{-1}{\frac{f}{2}-Qf_Q}\left[\frac{f}{2}+2\frac{\partial}{\partial t}\left[H_yf_Q\right]+6Hf_QH_y    \right]\label{34}\ee
and finally, (\ref{qqq1}), (\ref{qqq2}) and (\ref{qqq3}) give 
\be\delta= \frac{1}{\frac{f}{2}-Qf_Q}\left[\frac{\partial}{\partial t}\left[f_Q\left(H_y-H_x\right)\right]+3Hf_Q\left(H_y-H_x\right)   \right].\label{35}\ee
From(\ref{Q}), (\ref{H}) and (\ref{delta}) we can calculate 
\be 6H^2\left(1-\frac{\Delta}{2}\right)=-Q.\label{w1}\ee 
Using (\ref{w1}) in (\ref{qq1}) we obtain an expression for the energy density
\be \rho=\frac{f}{2}+6H^2f_Q-3H^2\Delta f_Q.\label{w2}\ee
This is vital for our discussion, which shows that for a given value of the mean Hubble parameter $H$ in the LRS-BI spacetime, the anisotropy of the expansion lowers down the energy density $\rho$; the highest possible energy density is achieved in case of isotropic expansion (i.e., $\Delta(t) = 0$) and the anisotropy of the expansion is constrained by the equation (\ref{w2}) and cannot vary arbitrarily. 

From (\ref{w2}), we can define the energy density associated with the anisotropy of the expansion by $\rho_{anis}=3H^2f_Q\Delta$ and write
\be
3H^2=\rho_{eff}=\rho+\rho_Q+\rho_{anis}
\ee
with $\rho_Q=3H^2-\frac{f}{2}-6H^2f_Q$. For $\Delta(t)\rightarrow 0$, other than vanishing $\rho_{anis}$, from (\ref{delta}) we obtain $A(t)\rightarrow B(t)$, so the expansion tends to be isotropic and we regain the usual isotropic and homogeneous FRW universe $A(t)=B(t)$ and by (\ref{35}), a vanishing $\delta$.

%%%%%%%%%%%%%%%%%%%%%%%%%%%%%%%%%%%%%%%%%%%%%%%%%%%%%%%%%%%%%%%%%%%%%%%%%%%%
%%%%%%%%%%%%%%%%%%%%%%%%%%%%%%%%%%%%%%%%%%%%%%%%%%%%%%%%%%%%%%%%%%%%%%%%%%%%
\section{Revisiting GR}\label{sec5}
We consider the case $f(Q)=Q$ to revisit the LRS-BI universe in GR. In this case the equations of motion (\ref{qq1})-(\ref{qq3}) reduces to
\bea
\rho&=&-\frac{Q}{2}\label{q1}\\
\omega\rho&=&-\frac{Q}{2}-2\left[\frac{\dot{A}}{A}\frac{\dot{B}}{B}+\left(\frac{\dot{B}}{B}\right)^2+\frac{\ddot{B}}{B} \right]\label{q2}\\
(\omega+\delta)\rho&=&-\frac{Q}{2}-\left[3\frac{\dot{A}}{A}\frac{\dot{B}}{B}+\left(\frac{\dot{B}}{B}\right)^2+\frac{\ddot{A}}{A}+\frac{\ddot{B}}{B} \right].\label{q3}
\eea

%Using (\ref{q1}), from (\ref{q2}) and (\ref{q3}) we obtain
%\bea(\omega-1)Q&=&4\left[\frac{\dot{B}}{B}\left(\frac{\dot{A}}{A}+\frac{\dot{B}}{B}\right)+\frac{\ddot{B}}{B}   \right]\\(\omega+\delta-1)Q&=&2\left[\frac{\dot{B}}{B}\left(\frac{\dot{B}}{B}+3\frac{\dot{A}}{A}\right)+\frac{\ddot{A}}{A}+\frac{\ddot{B}}{B}   \right]\eea

(\ref{w2}), in this case reduces to
\be \rho=3H^2\left(1-\frac{\Delta}{2}\right).\label{q4}\ee 
Therefore, $\Delta<2$ is essential to observe $\rho >0$ for a comoving observer, this result is equivalent to the generalized Friedmann equation for a Bianchi type-I spacetime (see \cite{ellis, barrow} for the generalized Friedmann equation).

This is a system of three equations in five unknowns $A,B,\rho,\omega$ and $\delta$. So to completely solve the system we need to impose two additional conditions. We start with the following scenario:

\subsection{The volumetric expansion law }\label{sub1}
As a popular choice in the literature concerning anisotropic models in GR \cite{rodrigues,akarsu}, we consider a constant $\rho_{eff}$ which produces a constant $H$ and thus the De Sitter volumetric expansion law \be V=c_1e^{3H_0t},\ee
where $c_1$ and $H_0$ are two positive constants. From (\ref{V}) we have
\be A(t)=\frac{c_1e^{3H_0t}}{B^2}.\label{44}\ee 
As the second condition to be imposed in the system, we consider two distinct cases as follows:

\subsubsection{Case I: $\omega=-1$}\label{sub11}
Now, using Equations \eqref{44} and \eqref{q1}-\eqref{q4}, we find out the following relations
\begin{equation}\label{a45}
B(t)=k_2 \sqrt[3]{e^{3 H_0 t}-k_1}
\end{equation}
and
\begin{equation}\label{a46}
A(t)=\frac{c_1 e^{3 H_0 t}}{k_2^2 \left(e^{3 H_0 t}-k_1\right)^{2/3}}.
\end{equation}
Using above two equations, we find the following expressions for energy density, $\delta$ and $\Delta$, respectively.

\begin{equation}
\rho=3 H_0^2 \left(1-\frac{k_1^2}{\left(e^{3 H_0 t}-k_1\right)^2}\right)
\end{equation}

\begin{equation}
\delta= -\frac{3 k_1^2}{e^{6 H_0 t}-2 k_1 e^{3 H_0 t}}
\end{equation}

\begin{equation}
\Delta =\frac{2 k_1^2}{\left(e^{3 H_0 t}-k_1\right)^2}.
\end{equation}

Also, it is observed that $\rho$ is positive throughout the evolution of the universe. The $\delta$ of the equation of state parameter monotonically decreases as t tends to infinite. This behavior of skewness parameter suggests that the anisotropic expansion is high during the early phase of the universe and later reduces to isotropic expansion. For this case, the anisotropic expansion takes its values less than 2 and reduces to zero as t tends to be infinite. These results suggest that the anisotropic expansion of our universe is high during the early time, and later, it becomes isotropic.

\subsubsection{Case II: $\rho$ = constant $(\gamma)$.}\label{sub12}

From Equations \eqref{44} and \eqref{q1}-\eqref{q4}, we find the following relations for anisotropic scale factors

\begin{equation}\label{50}
B(t)=k_3 e^{\left(H_0\pm \frac{\sqrt{9 H_0^2-3 \gamma}}{3}\right)t}
\end{equation}

\begin{equation}\label{51}
A(t)=\frac{c_1}{k_3^2} e^{\left(H_0\mp \frac{2\sqrt{9 H_0^2-3 \gamma}}{3}\right)t}.
\end{equation}
Now, using \eqref{50} and \eqref{51}, we can write $(\delta)$ and $(\omega)$ as 
\begin{equation}
\delta= 2+\frac{-6 H_0^2+ H_0\sqrt{9H_0^2-3\gamma}}{\gamma},\,\,\,\text{or, } 2-\frac{6 H_0^2+ H_0\sqrt{9H_0^2-3\gamma}}{\gamma},
\end{equation}

\begin{equation}
\omega =\frac{\sqrt{9 H_0^2-3 \gamma}+3 H_0}{\sqrt{9 H_0^2-3  \gamma}-3 H_0} ,\,\, \text{or, }
\frac{\sqrt{9 H_0^2-3  \gamma}-3 H_0}{\sqrt{9 H_0^2-3  \gamma}+3 H_0},
\end{equation}
and, the anisotropy measures as
\begin{equation}
\Delta = 2-\frac{2 \gamma}{3 H_0^2}.
\end{equation}

With the consideration of constant energy density $\rho$ in this case, we derive the solutions for anisotropic scale factors. From all of the above expressions, it is clear that $\gamma \leq 3 H_0^2$. For $\gamma = 3 H_0^2$, our model shows the isotropic expansion of the universe, because $\delta=0$, $\Delta=0$, and $\omega=-1$. However, when  $\gamma < 3 H_0^2$, then we have two cases such as $\omega <-1$, or $\omega >-1$ and $\Delta>0$. Also, it can be observed that while the pressure along the x-axis shows the phantom behavior, i.e., $\omega<-1$ and quintessence behavior i.e., $\omega>-1$, the pressure along y and z-axes always shows the phantom behavior, i.e., $\omega+\delta<-1$. We summarize the results in below table (there, first pair of scale factors represents $A(t)=\frac{c_1}{k_3} e^{\left(H_0- \frac{2\sqrt{9 H_0^2-3 \gamma}}{3}\right)t},\,\ B(t)=k_3 e^{\left(H_0+ \frac{\sqrt{9 H_0^2-3 \gamma}}{3}\right)t}$ and second pair represents the other pair of scale factors with $H_0=1$. As a result, $\rho=\gamma$ should lie in $(0,3]$).

\begin{center}

Table-I
\begin{tabular}{|cc|c|c|c|c}

%\cline{3-6}
%& & \multicolumn{4}{ c| }{Primes} \\ 
\cline{1-5}
& & $\omega$ & $\delta$ & $\Delta$  \\ \cline{1-5}
\multicolumn{1}{ |c | }{\multirow{2}{*}{First pair} } &
\multicolumn{1}{ |c| }{$\gamma\rightarrow 0$} & Phantom region ($\omega<-1$) & $-\infty$ & 2 &     \\ \cline{2-5}
\multicolumn{1}{ |c  }{}                        &
\multicolumn{1}{ |c| }{$\gamma\rightarrow 3$} & $\Lambda$CDM ($\omega\simeq -1$) & 0 & 0 &      \\ \cline{1-5}
\multicolumn{1}{ |c  }{\multirow{2}{*}{Second pair} } &
\multicolumn{1}{ |c| }{$\gamma\rightarrow 0$} & Quintessence ($\omega < -1$) & $-\infty$ & 2 &  \\ \cline{2-5}
\multicolumn{1}{ |c  }{}                        &
\multicolumn{1}{ |c| }{$\gamma\rightarrow 3$} & $\Lambda$CDM ($\omega\simeq -1$) & 0 & 0 &   \\ \cline{1-5}

\end{tabular}
%%%%%%%%%%%%%%%%%%%%%%%%%%%%%%%%%%%%%%%%%%%%%%%%%%%%%%%%%%%%%%%%%%%%%%%%%%%
%%%%%%%%%%%%%%%%%%%%%%%%%%%%%%%%%%%%%%%%%%%%%%%%%%%%%%%%%%%%%%%%%%%%%%%%%%%
\end{center}

\subsection{Anisotropic relation $(\theta^2 \propto \sigma^2) $}\label{sub2}
As the second set of additional conditions, we first consider a physical condition that the shear is proportional to the expansion scalar and this leads to the relation 
\be A=B^n,\label{55}\ee
where $n(\neq  0,\, 1)$ is an arbitrary real number. This physical law is imposed on the basis of the observations of the velocity redshift relation for extragalactic sources which suggest that the Hubble expansion of the universe may achieve isotropy when $\frac{\sigma}{\theta}$ is constant \cite{a82}. The condition was used in several occasions in the literature \cite{rodrigues, bishi, sahoo}.

As earlier, we consider two distinct cases to close the system and compare them.

\subsubsection{Case I: $\omega=-1$.}\label{sub21}

Using equation \eqref{qqq1}, \eqref{34}, \eqref{35}, and \eqref{55}, we find the following expressions for scale factors
\begin{equation}\label{a56}
A(t)=\left[(1-n) \left(k_4 t+k_5\right)\right]{}^{\frac{n}{1-n}}
\end{equation}
\begin{equation}\label{a57}
B(t)=\left[(1-n) \left(k_4 t+k_5\right)\right]{}^{\frac{1}{1-n}}.
\end{equation}
 
Following the above anisotropic scale factors, we can express
\begin{equation}
\rho=\frac{k_4^2 (2 n+1)}{(n-1)^2 \left(k_4 t+k_5\right){}^2}
\end{equation}

\begin{equation}
\delta = 1-n
\end{equation}

\begin{equation}
\Delta = \frac{2 (n-1)^2}{(n+2)^2}.
\end{equation}

\subsubsection{Case II: $\rho$ = constant $(\beta)$.}\label{sub22}

Using equation \eqref{qqq1}, \eqref{34}, \eqref{35}, and \eqref{55}, we find the following expressions for scale factors

\begin{equation}\label{a61}
A(t) = \left[k_6 e^{t \sqrt{\frac{\beta}{2 n+1}}}\right]^n
\end{equation}
\begin{equation}\label{a62}
B(t)= k_6 e^{t \sqrt{\frac{\beta}{2 n+1}}}.
\end{equation}

Now, we can write the cosmological parameters as follows

\begin{equation}
\delta = -\frac{(n-1) n}{2 n+1}
\end{equation}

\begin{equation}
\omega = -\frac{3}{2 n+1}
\end{equation}

\begin{equation}
\Delta =\frac{2 (n-1)^2}{(n+2)^2}.
\end{equation}

%%%%%%%%%%%%%%%%%%%%%%%%%%%%%%%%%%%%%%%%%%%%%%%%%%%%%%%%%%%%%%%%%%%%%%%%%%%
%%%%%%%%%%%%%%%%%%%%%%%%%%%%%%%%%%%%%%%%%%%%%%%%%%%%%%%%%%%%%%%%%%%%%%%%%%%

\section{Analysing $f(Q)$-models}\label{sec6}
In this section, we consider that the anisotropy in the universe is given by the same forms of scale factors $A(t)$ and $B(t)$ as in the previous section. We use the four sets of values of $A(t)$ and $B(t)$ obtained in Section \ref{sub1} and Section \ref{sub2} and investigate the main goal of the present study, the dynamics in particular $f(Q)$-models. This comparative analysis illustrates the impact of the non-metricity in terms of the modifications in the matter content of the universe through the new terms collected across $f(Q)$ in the motion equations. For our present discussion, we consider the well-analysed model $f(Q)=Q+\alpha Q^2$. The polynomial forms of $f(Q)$ have been studied widely. For instance, Mandal et al. \cite{Mandal/2020b} studied the energy conditions to examine the viability of the cosmological models with the observational measurements of cosmographic parameters. Hasan et al. \cite{Q3} studied the various types of wormhole models, Jim\'enez et al. \cite{jim} explored the cosmological perturbation scenario, bouncing scenarios to avoid the initial singularity problem studied in \cite{man}, and dynamical analysis of the cosmological model analyzed in \cite{khy}.  Therefore, it is worthy to consider a polynomial form of $f(Q)$ to study the Bianchi universe.

We start with the first condition:\\

\subsection{Case I: $\rho_{eff}$ is constant and the EoS $\omega=-1$. }

We begin with the expression of $A(t)$ and $B(t)$ given respectively by (\ref{a45}) and (\ref{a46}) and obtain the energy density as

\begin{equation}
\rho =3 H_0^2 \left(-18 \alpha  H_0^2-\frac{18 \alpha  H_0^2 k_1^4}{\left(e^{3 H_0 t}-k_1\right)^4}+\frac{k_1^2 \left(36 \alpha  H_0^2-1\right)}{\left(e^{3 H_0 t}-k_1\right)^2}+1\right).
\end{equation}
As $t\rightarrow \infty$, the energy density $\rho\rightarrow 3H_0^2(1-18\alpha H_0^2)$. Therefore, for a positive energy density throughout the evolution history we must make sure that the parameter $\alpha$ satisfies $\alpha<\frac{1}{18H_0^2}$ for the present case.

The skewness parameter $\delta$ can be calculated as
\begin{equation}
\delta = \frac{3 k_1^2 e^{-3 H_0 t} \left(2 k_1 \left(24 \alpha  H_0^2-1\right) e^{3 H_0 t}+\left(1-12 \alpha  H_0^2\right) e^{6 H_0 t}+k_1^2\right)}{\left(e^{3 H_0 t}-2 k_1\right) \left(-2 k_1 \left(18 \alpha  H_0^2-1\right) e^{3 H_0 t}+\left(18 \alpha  H_0^2-1\right) e^{6 H_0 t}-k_1^2\right)}
\end{equation}
which vanishes as $t\rightarrow \infty$ since $18\alpha H_0^2\neq 1$, contributing to isotropization of the fluid. Finally, the anisotropy of the universe is given by
\begin{equation}
\Delta = \frac{2 k_1^2}{\left(e^{3 H_0 t}-k_1\right)^2}
\end{equation}
which also clearly fades away as time increases, thus entering into the isotropic Robertson-Walker geometry.

\subsection{ Case II: $\rho_{eff}$ is constant and $\rho$ is constant.}
 
In this case, we assume the expressions of $A(t)$ and $B(t)$ are given respectively by (\ref{50}) and (\ref{51}). Following the same procedure as above, we calculate the parameters as

\begin{align}\label{69}
\omega = \frac{(2 \alpha  \gamma-1) \sqrt{9 H_0^2-3 \gamma}+3 H_0 (6 \alpha  \gamma-1)}{(6 \alpha  \gamma-1) \left(\sqrt{9 H_0^2-3 \gamma}-3 H_0\right)},\,\, \text{or,} \frac{(2 \alpha  \gamma-1) \sqrt{9 H_0^2-3 \gamma}+H_0 (3-18 \alpha  \gamma)}{(6 \alpha  \gamma-1) \left(\sqrt{9 H_0^2-3 \gamma}+3 H_0\right)}
\end{align}

\begin{align}\label{70}
\delta = \frac{3 H_0 (4 \alpha  \gamma-1) \sqrt{9 H_0^2-3 \gamma}}{\gamma (6 \alpha  \gamma-1)},\,\,\ \text{or,} \frac{3 H_0 (4 \alpha  \gamma-1) \sqrt{9 H_0^2-3 \gamma}}{\gamma (1-6 \alpha  \gamma)}.
\end{align}
%\begin{equation}
%\Delta= 2-\frac{2 \text{C2}}{3 \text{H0}^2}
%\end{equation}
%%%%%%%%%%%%%%%%%%%%%%%%%%%%%%%%%%%%%%%%%%%%%%%%%%%%%%%%%%%%%%%%%%%%%%%%%%%
%%%%%%%%%%%%%%%%%%%%%%%%%%%%%%%%%%%%%%%%%%%%%%%%%%%%%%%%%%%%%%%%%%%%%%%%%%%
For this case, we present two sets of parameters as we have two pairs of scale factors. Previously, we have discussed about the other parameters except $\alpha$ in equation \eqref{69} and \eqref{70}. Here, we observe that for any values of $\alpha$ with the values of parameters in the subsection \ref{sub1}, $\delta$ lies in its positive range and $\omega$ represents phantom region for first pair of solution and for second pair of solution $\delta$ lies in phantom region and $\omega$ lies in quintessence region.

\subsection{Case-III: $\theta^2 \propto \sigma^2 $ and the EoS $\omega=-1$.}

In this section, we consider the expressions of $A(t)$ and $B(t)$ given in (\ref{a56}) and (\ref{a57}). For this anisotropic scale factors, we find the following expressions for energy density $\rho$ and skewness parameter $\delta$

\begin{equation}
\rho = \frac{k_4^2 (2 n+1) \left((n-1)^2 \left(k_4 t+k_5\right){}^2-6 \alpha  k_4^2 (2 n+1)\right)}{(n-1)^4 \left(k_4 t+k_5\right){}^4}
\end{equation}
%\begin{equation}\omega = -1+\frac{16 k_4^2 \alpha  (n-1)}{(n-1)^2 (k_4 t+k_5)^2-6 k_4^2 \alpha  (2 n+1)}\end{equation}
\begin{equation}
\delta =\frac{4 k_4^2 \alpha  (n-1) (4 n-1)-(n-1)^3 (k_4 t+k_5)^2}{(n-1)^2 (k_4 t+k_5)^2-6 k_4^2 \alpha  (2 n+1)}.
\end{equation}
As $t$ is increasing, $\delta\rightarrow 1-n$. Hence, the vanishing skewness parameter requirement produces $n=1$. In other words, in the present case isotropization of the gravitational fluid pushes the anisotropy of the universe to enter into the isotropic Robertson-Walker geometry by providing $A=B$. It is observed that the energy density reduces as time goes on and equation of state converges towards $\Lambda$CDM with time. The skewness parameter takes its value in positive range.

\subsection{Case-IV: $\theta^2 \propto \sigma^2 $ and $\rho$ is constant.}

In this subsection, we presume the expressions (\ref{a61}) and (\ref{a62}). Using those expressions, we obtain

\begin{equation}
\omega = \frac{3-2 \alpha  \beta (2 n+7)}{(2 n+1) \left(6 \alpha  \beta-1\right)}
\end{equation}

\begin{equation}
\delta = -\frac{(n-1) (n+2) \left(4 \alpha  \beta-1\right)}{(2 n+1) \left(6 \alpha  \beta-1\right)}
\end{equation}

Here, we observe that the values of the parameters as constant. Let's take the values of parameters from the previous subsection \ref{sub2}, then for any values of $\alpha$. $ \omega$ and $\delta$ behaves oppositely (i.e., if $\omega$ behaves like phantom, then $\delta$ behaves like quintessence and vice-versa.)

\section{Conclusion}\label{sec7}

In this manuscript, we have focused on exploring the Bianchi Universe in a novel modified $f(Q)$-gravity theory, where the gravitational interaction is demonstrated without the presence of curvature and torsion. Here, we have derived the motion equations for $f(Q)$-gravity in the anisotropic yet homogeneous LRS Bianchi-I spacetime in the presence of a single anisotropic perfect fluid with a dynamic equation of state (EoS) parameter and energy density. We have proved the conservation of energy-momentum from contracted Bianchi identity and thus secured a non-constant energy density even for DE EoS parameter $\omega =-1$. In order to re-obtain the anisotropic dark energy results as an analogy to the GR, we have first considered $f(Q)=Q$ in the section \ref{sec5}. Four different physical scenarios have been discussed, two cases under the assumption of volumetric expansion law and the other two for anisotropic relation. For all the four cases, we have found the expressions for the anisotropic scale factors $A(t)$ and $B(t)$. 
To proceed further in our study of isotropization history in the cosmological evolution of universe in the $f(Q)$-theory, we have considered a widely studied polynomial form of $f(Q)=Q+\alpha Q^2$. 

To investigate this particular $f(Q)$ model, we have made use of the expressions of the four pairs of anisotropic scale factors $A(t)$ and $B(t)$ from the previous section \ref{sec5}. Such way we have obtained modifications in the matter content of the universe (the anisotropy of the fluid and the new contribution of the nonlinear terms of $Q$). We have derived the energy density $\rho$, skewness parameter $\delta$, and anisotropic parameter $\Delta$. Our investigation has illustrated that at the early time, the universe presented unequal scale factor in the $x$ and $y$ directions, and as the late-time is reached, they display equality, and in due process remove the anisotropy in the gravitational fluid. Furthermore, we have analyzed the energy density $\rho$, EoS $\omega$, the anisotropic parameter $\Delta$, skewness parameter $\delta$ for four separate sets of physical conditions in detail for each case. Our study has concluded that even in the presence of an anisotropic fluid, the universe could approach isotropy and isotropizes the anisotropy of the fluid at the same time. So we cannot rule out a priori the possibility of anisotropic DE, even though the present universe shows an isotropic expansion.  

Moreover, nine-year Wilkinson Microwave Anisotropy probe (WAMP) observations suggest small deviations from the isotropy \cite{wamp9}, which motivated us to study the Bianchi universe as it could lead to more realistic results. Also, the observational constraint on the DE EoS $\omega$ such as $\omega=-1.10\pm 0.14$ \cite{wamp7}, $-0.14<1+\omega <0.12$ \cite{wmap2} suggest, its value should be very close to $-1$. In addition, several observational studies on Bianchi identities in the gravitational frameworks show the same behavior of $\omega$ to present the late-time acceleration of the universe \cite{A1,A2,A4,A5}. Although we have theoretically constructed and studied the anisotropic universe in a modified gravity framework, our outcomes align with the observational results. In further studies, it would be interesting to test these models against the observational data such as Hubble, Pantheon, BAO. Some of these tests will be addressed in the near future, and we hope to report them.

\section*{Acknowledgments}

A.D. and L.T.H. are supported in part by the FRGS research grant (Grant No. FRGS/1/2021/STG06/UTAR/02/1). S.M. acknowledges Department of Science \& Technology (DST), Govt. of India, New Delhi, for awarding Senior Research Fellowship (File No. DST/INSPIRE Fellowship/2018/IF180676).

%%%%%%%%%%%%%%%%%%%%%%%%%%%%%%%%%%%%%%%%%%%%%%%%%%%%%%%%%%%%%%%%%%%%%%%%%%%
%%%%%%%%%%%%%%%%%%%%%%%%%%%%%%%%%%%%%%%%%%%%%%%%%%%%%%%%%%%%%%%%%%%%%%%%%%%

\section{Appendix}

The non-vanishing Levi-Civita connections of (\ref{metric}) are:
\begin{align*} 
\mathring{\Gamma}^t{}_{xx} &= A\dot{A} \,, \quad \mathring{\Gamma}^t{}_{yy} = \mathring{\Gamma}^t{}_{zz} = B\dot{B}  \\
\mathring{\Gamma}^x{}_{tx} &= \frac{\dot{A}}{A} = \mathring{\Gamma}^x{}_{xt} \,, \quad \mathring{\Gamma}^y{}_{ty} = \mathring{\Gamma}^z{}_{tz} = \frac{\dot{B}}{B} = \mathring{\Gamma}^y{}_{yt} = \mathring{\Gamma}^z{}_{zt} \,.
\end{align*}

The corresponding Einstein tensors are: 
\begin{align*}
\m{G}_{tt} &= 2\frac{\dot{A}}{A} \frac{\dot{B}}{B} + \left(\frac{\dot{B}}{B} \right)^2  \\
\m{G}_{xx} &= -A^2 \left( \frac{\dot{B}}{B} \right)^2 -2 A^2 \frac{\ddot{B}}{B} \\
\m{G}_{yy} &= -\frac{\ddot{A}}{A} B^2 -\frac{\dot{A}}{A} B\dot{B} - B \ddot{B} \\
\m{G}_{zz} &= \m{G}_{yy} \, .
\end{align*}
The non-vanishing non-metricity tensors are:
\begin{align*}
Q_{txx} = 2A\dot{A};\quad Q_{tyy} = 2B\dot{B}; \quad Q_{tzz} = 2B\dot{B}.
\end{align*}
The superpotential tensors are:
\begin{align*}
P^{txx} &= - \frac{\dot{B}}{A^2 B} \\
P^{tyy} &= -\frac{1}{2}\frac{\dot{A}}{A B^2} -\frac{1}{2}\frac{\dot{B}}{B^3} \\
P^{tzz} &= P^{tyy} \\
P^{xxt} &= -\frac{1}{4}\frac{\dot{A}}{A^3} + \frac{1}{2}\frac{\dot{B}}{A^2 B} =P^{xtx} \\
P^{yyt} &= \frac{1}{4}\frac{\dot{A}}{AB^2} = P^{yty} \\
P^{zzt} &= \frac{1}{4}\frac{\dot{A}}{AB^2} = P^{ztz}.
\end{align*}

%%%%%%%%%%%%%%%%%%%%%%%%%%%%%%%%%%%%%%%%%%%%%%%%%%%%%%%%%%%%%%%%%%%%%%%%%%%
%%%%%%%%%%%%%%%%%%%%%%%%%%%%%%%%%%%%%%%%%%%%%%%%%%%%%%%%%%%%%%%%%%%%%%%%%%%

\end{document}